\documentclass[aps,prl,twocolumn,groupedaddress,superscriptaddress,showpacs]{revtex4-1}

\usepackage{graphicx}
\usepackage{color}
\usepackage{amsmath}
\usepackage{amsfonts}
\usepackage{amssymb}
\usepackage{dcolumn}
\usepackage{hyperref}
\hypersetup{colorlinks=true,urlcolor=blue,linkcolor=blue,citecolor=blue}
\hfuzz 1pt \vfuzz 1pt \definecolor{orange}{rgb}{1,0.5,0}
\definecolor{green}{rgb}{0,0.5,0}
\usepackage[final]{pdfpages}

\begin{document}
\title{Critical fragmentation properties of random drilling: How many
random holes need to be drilled to collapse a wooden cube?}
%
%
%
\author{K.~J.~Schrenk} \email{kjs73@cam.ac.uk}
\affiliation{Computational Physics for Engineering Materials, IfB, ETH
Zurich, Wolfgang-Pauli-Strasse 27, CH-8093 Zurich, Switzerland}
\affiliation{Department of Chemistry, University of Cambridge, Lensfield
Road, Cambridge, CB2 1EW, U.K.}
\author{M.~R.~Hil\'ario} \email{mhilario@mat.ufmg.br}
\affiliation{Departamento de Matem\'atica, Universidade Federal de Minas
Gerais, Av.~Antonio Carlos, 6627
- PO Box 702 - 30161-970, Belo Horizonte, MG, Brazil}
  \affiliation{Section de Math\'ematiques, Universit\'e de Gen\`eve, 2-4
Rue du Li\`evre, 1211, Gen\`eve, Switzerland}
\author{V.~Sidoravicius} \email{vladas@impa.br} 
\affiliation{Courant Institute of Mathematical Sciences, New York
University, 251 Mercer Street, New York, NY 10012, US}
\affiliation{New York University -- Shanghai, 1555 Century Avenue,
Pudong New Area, Shanghai, 200122, China}
\affiliation{CEMADEN, Avenida Doutor Altino Bondensan, 500, S\~ao Jos\'e
dos Campos, SP, 12247-016, Brazil}
\author{N.~A.~M.~Ara\'ujo} \email{nmaraujo@fc.ul.pt}
\affiliation{Departamento de F\'{\i}sica, Faculdade de Ci\^{e}ncias,
Universidade de Lisboa, 1749-016 Lisboa, Portugal, and Centro de
F\'isica Te\'orica e Computacional, Universidade de Lisboa, 1749-016
Lisboa, Portugal}
\author{H.~J.~Herrmann} \email{hans@ifb.baug.ethz.ch}
\affiliation{Computational Physics for Engineering Materials, IfB, ETH
Zurich, Wolfgang-Pauli-Strasse 27, CH-8093 Zurich, Switzerland}
\affiliation{Departamento de F\'{\i}sica, Universidade Federal do
Cear\'a, 60451-970, Fortaleza, CE, Brazil}
\author{M.~Thielmann} \email{marcel.thielmann@uni-bayreuth.de}
\affiliation{Bayerisches Geoinstitut, University of Bayreuth,
Universit\"{a}tsstra\ss e 30, 95440 Bayreuth, Germany}
\author{A.~Teixeira} \email{augusto@impa.br} \affiliation{Instituto
Nacional de Matem\'atica Pura e Aplicada, Est.~Dona Castorina, 110,
22460-320, Rio de Janeiro, RJ, Brazil}
%
%
%
\pacs{02.50.Cw, 89.75.Da, 64.60.ah}
%
\begin{abstract}
A solid wooden cube fragments into pieces as we sequentially drill
holes through it randomly. This seemingly straightforward observation
encompasses deep and nontrivial geometrical and probabilistic behavior
that is discussed here. Combining numerical simulations and rigorous
results, we find off-critical scale-free behavior and a continuous
transition at a critical density of holes that significantly differs
from classical percolation.
\end{abstract}
%
\maketitle
%

The connectivity of a solid block of material strongly depends
on the density of defects. To systematically study this dependence one
must find an experimental way to create defects inside the solid. 
For example, in 2D one can simply punch holes in a
sheet and measure the physical properties of the remaining
material. But in 3D, inducing localized defects is not simple.
One conventional solution consists in perforating the material by
drilling holes or laser ablation from the surface~\cite{Simon96,Nolte97}. 

\begin{figure}
\begin{center}
  \includegraphics[width=\columnwidth]{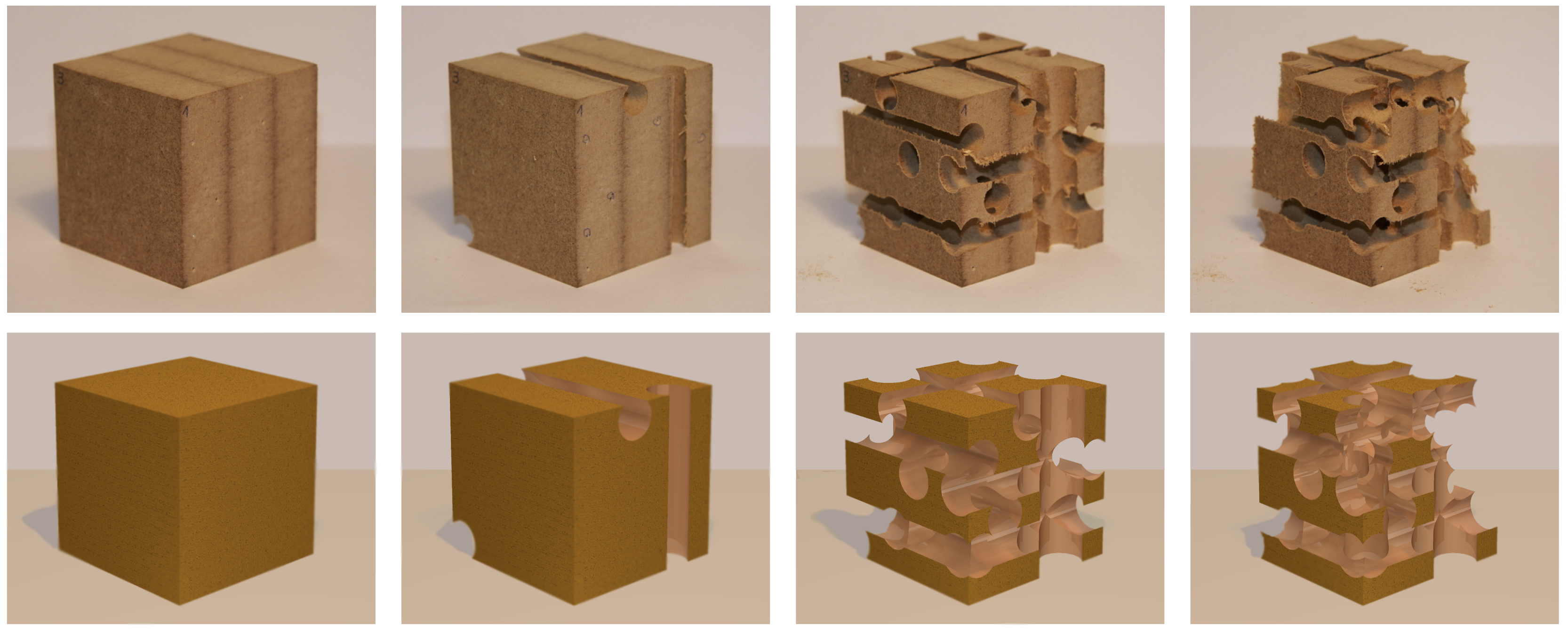}

\hspace{0.75cm} (a) \hspace{1.5cm} (b) \hspace{1.5cm} (c) \hspace{1.5cm} (d) \hspace{0.75cm}

\end{center} \caption{ \label{fig::zj_images_experiment_rendering}
(color online) Cube drilling.  The upper row shows photos of the
experimental setup; the lower panels are the corresponding numerical
results.  The faces of the cube are divided with a square-lattice mesh
of linear size $L=6$, such that each face can be drilled $L^2=36$ times.
From left to right, (a) -- (d), the number of drilled holes per face are
(a) 0, (b) 1, (c) 6, and (d) 8.  From our numerical results for the
position of the transition in the thermodynamic limit (see main text),
one estimates that around $13$ holes need to be drilled for the cube to
disconnect.  } \end{figure}

In a table-top experiment, we start with a solid cube of wood and plot
on each face a square-lattice mesh of $L$ by $L$ cells.  Initially, the
cube has no holes.  Sequentially, for each one of three perpendicular
faces, we randomly choose one square-cell and drill a hole having a
radius of $1/\sqrt{2}$ cell lengths to the other side of the cube. 
We repeat this process iteratively until the entire 
structure collapses into small pieces and the bottom and top part 
of the cube are no longer connected.
The first row of Fig.~\ref{fig::zj_images_experiment_rendering}(a)-(d)
shows the result of the drilling process of a real cube with edge length
6 cm (manufactured from 2 cm thick plates of medium-density fiberboard
(MDF)), where holes were drilled with a diameter of 1 cm. As the
drilling proceeds, pieces get disconnected and eventually the entire
structure collapses. 

%
%
%
Numerically, we start with a three-dimensional cubic lattice of $L^3$
sites and fix three perpendicular faces. A fraction $1-p$ square-cells on each face
is randomly selected and all sites along
the line perpendicular to that face are removed (second row of
Fig.~\ref{fig::zj_images_experiment_rendering}).
Thirty years ago, Y.
Kantor~\cite{Kantor86} numerically studied this model on lattices of 
up to $10^6$ sites and concluded that the critical fragmentation properties
of this model are in the same universality class as random
percolation~\cite{Stauffer79,Stauffer94,Grimmett99}. Here, we combine
rigorous results and large-scale numerical simulations, considering lattices three
orders of magnitude larger in size, to show that this
is not the case. Removing entire rows at once induces strong long-range
directional correlations and the critical behavior departs from random
percolation.  Also remarkably, while in random fragmentation power-law
scaling is solely observed around the critical threshold, here we find
it in an entire off-critical region.
These findings suggest that long-range directional correlations
lead to a rich spectrum of critical phenomena which need to be understood. 
Possible implications for other complex percolation models are discussed 
in the conclusions.

\textit{Threshold.} The average total number of drilled holes is 
$3(1-p)L^2$ and the asymptotic probability that a site in the bulk is
not removed is $p^3$.  
We first measure the threshold $p_c$ at which the
cube collapses for different lattice sizes, up to $L=1024$, using different estimators
of the transition point, as discussed in the Supplemental
Material~\cite{SM}.
Extrapolating the data to the limit $L\to\infty$,
gives $p_c=0.6339\pm0.0005$, consistent with the value estimated by
Kantor using Monte Carlo renormalization group
techniques~\cite{Kantor86} (see Supplemental Material~\cite{SM}). This
threshold is larger than the two-dimensional square-lattice percolation
threshold ($p_{2D}$)~\cite{Ziff11b, Jacobsen14} and smaller than the
cubic root of the one for the three-dimensional simple cubic
lattice~\cite{Wang13}.  

\begin{figure}
\begin{tabular}{l}
(a) \\
\includegraphics[width=\columnwidth]{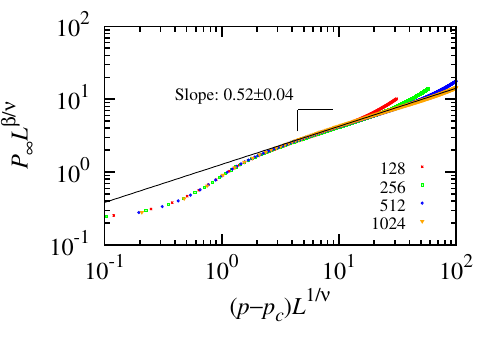} \\
(b) \\
\includegraphics[width=\columnwidth]{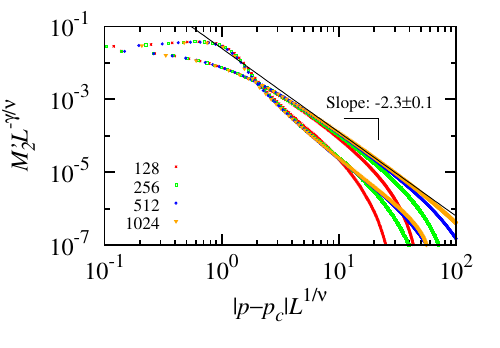} \\
\end{tabular}
\caption{
\label{fig::log_log_fss_plots}
(color online) (a) Double-logarithmic plot of the rescaled order
parameter $P_\infty L^{\beta/\nu}$ as function of the scaling variable
$(p-p_c)L^{1/\nu}$ for different lattice sizes $L$.  The linear part has
a slope of $\beta=0.52\pm0.04$, consistent with $\beta/\nu$ being the
same as for three-dimensional percolation, but with a different exponent
$1/\nu=0.92$.  The present value of $\beta$ is different from the
two-dimensional one $\beta=5/36\approx0.139$ \cite{Stauffer94,
Smirnov01b} and the one in three dimensions, $\beta\approx0.417$
\cite{Deng05b}.  (b) Double-logarithmic plot of the rescaled second
moment $M_2'L^{-\gamma/\nu}$, with $\gamma/\nu=2.0452$, as function of
the scaling variable $|p-p_c|L^{1/\nu}$, with $1/\nu=0.92$, for
different lattice sizes $L$.  The solid black line is a guide to the eye
with slope $-2.3$. The two sets of data points correspond to
the sub-critical ($p<p_c$) and super-critical ($p>p_c$) regions.
}
\end{figure}

\textit{Static exponents.} We consider the fraction $P_\infty$ of 
sites in the largest cluster of connected sites (see Supplemental
Material for more data of $P_\infty(p)$ and of other
observables~\cite{SM}).  $P_\infty$ is the standard order parameter in
percolation identifying the transition from a disconnected to a globally
connected state.  For the drilling model, the situation will turn out to
be more complicated.  
Figure~\ref{fig::log_log_fss_plots}(a) shows a
double-logarithmic plot of the order parameter, rescaled by a power of
the lattice size $P_\infty L^{\beta/\nu}$ as function of the distance to
the transition $\lvert p-p_c\rvert L^{1/\nu}$.  Based on
finite-size scaling analysis \cite{Stauffer94}, we find that the critical exponent 
of the order parameter is $\beta =
0.52\pm0.04$, and the inverse of the correlation length
exponent is $1/\nu = 0.92\pm0.01$ (see Supplemental Material).
We note that both $\beta$ and $\nu$
are different from the corresponding values for 2D and 3D classical
percolation. However, somehow surprisingly, the exponent ratio
$\beta/\nu=0.50\pm0.04$ is within error bars the same as for 3D
percolation.  Thus, while the fractal dimension of the largest cluster
(given by $d_f=d-\beta/\nu$) is consistent with the one for 3D
percolation, the larger value of $\beta$ (compared to 3D percolation)
implies that the transition from the connected to the disconnected state
is less abrupt (see also Supplemental Material).  We consider next the
behavior of the second moment of the cluster size distribution $M_2'$,
excluding the contribution of the largest cluster.  As shown in
Fig.~\ref{fig::log_log_fss_plots}(b), the finite-size scaling analysis
gives that the susceptibility critical exponent is $\gamma = 2.3 \pm
0.1$.  We note that our results for the static critical exponents are
within error bars consistent with the scaling relation $2\beta + \gamma
= d \nu$ (for $d=3$).

\begin{figure}
\includegraphics[width=\columnwidth]{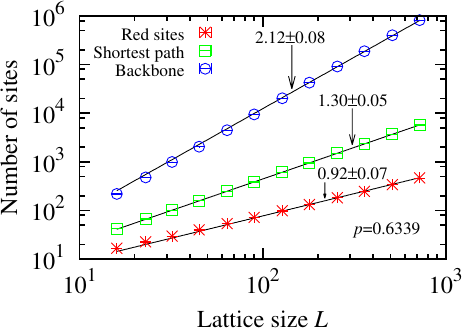}
\caption{
\label{fig::backbone_loglog}
(color online) Number of sites in the backbone of the spanning cluster,
length of its shortest path (chemical distance), and red sites in the
backbone, measured at $p=p_c=0.6339$, as function of the lattice
size $L$. Considering the local slopes of the data, we obtain the
following fractal dimensions: ${d_\mathrm{BB}={2.12\pm0.08}}$,
${d_\mathrm{SP}={1.30\pm0.05}}$, and ${d_\mathrm{RS}={0.92\pm0.07}}$.
The fractal dimension of the red sites is, within error bars,
compatible with the value ${1/\nu\approx0.915}$, found from the finite
size scaling behavior of the $p_c$ estimators (see
Fig.~S1), and the relation $d_\mathrm{RS}=1/\nu$
\cite{Coniglio89}. Here, $d_\mathrm{BB}$ is larger than in classical
three-dimensional percolation, where ${d_\mathrm{BB}=1.875\pm0.003}$
\cite{Herrmann84, Rintoul94, Deng04d}, while $d_\mathrm{SP}$ is smaller
than the classical value ${d_\mathrm{SP}=1.3756\pm0.0006}$
\cite{Zhou12}. The solid lines are guides to the eye. To extract the
fractal dimensions, we analyzed the local slopes as proposed
in Ref.~\cite{Grassberger99}. Results are averages over at least
${5\times10^3}$ samples.
}
\end{figure}

\textit{Dynamical exponents.} The transport properties of the largest
cluster at the critical threshold, $p_c$,  are intimately related to
dynamical critical exponents and they can be measured by quantifying
three sets of sites in the largest cluster \cite{Herrmann84}.
First, we consider the so-called red sites. A site is considered a red
site if its removal would lead to the collapse of the largest cluster
\footnote{Numerically, we consider those sites that if removed will
disconnect all possible paths between a unit site in the top and one in
the bottom faces of the cube. The sites are selected such that their
Euclidean distance is maximized.}. The red sites form a fractal set of
fractal dimension $d_\mathrm{RS}=0.92\pm0.07$ (see
Fig.~\ref{fig::backbone_loglog}), which is compatible with the inverse
of the correlation length exponent $\nu$ that we obtained from the
finite-size scaling analysis in Fig.~\ref{fig::log_log_fss_plots},
$d_\mathrm{RS}=1/\nu$, as predicted by Coniglio for classical
percolation~\cite{Coniglio89}. However the value of $d_\mathrm{RS} =
1/\nu$ for the drilling transition is very different from the classical
3D percolation result $1/\nu=1.1437\pm0.0006$~\cite{Deng05b}.
Figure~\ref{fig::backbone_loglog} also shows that the shortest path 
connecting the top and bottom sides of the largest cluster is
a fractal of fractal dimension $d_\mathrm{SP} = 1.30\pm0.05$.  Finally,
the backbone of the largest cluster between its bottom and top ends is
defined as the set of sites that would carry current if a potential
difference is applied between the cluster ends (also known as
bi-connected component). The backbone fractal dimension is determined
as $d_\mathrm{BB}=2.12\pm0.08$, which is larger than in classical 3D
percolation, where ${d_\mathrm{BB}=1.875\pm0.003}$~\cite{Herrmann84,
Rintoul94, Deng04d}.  Qualitatively, an increase in the backbone fractal
dimension is compatible with a simultaneous decrease in the shortest
path fractal dimension, since both correspond to a more compact
backbone, similar to what is observed in long-range correlated
percolation~\cite{Prakash92, Schrenk13b}. Thus, although the fractal
dimension of the largest cluster is similar in both classical
percolation and drilling, the internal structure of the largest cluster
is significantly different. This implies that transport and mechanical
properties of the largest cluster follow a different scaling.
%


\textit{Cluster shape.} Given the highly directional nature of the
drilling process, we analyze the symmetry of the different clusters. In
particular, we consider them as rigid bodies, consisting of
occupied sites at fixed relative positions, and look at the eigenvalues
and eigenvectors of their inertia tensors \cite{Rudnick87,
Mansfield13}. The numerical results show that, when compared to
classical percolation clusters, the drilling transition clusters are
more anisotropic, their orientations being mainly aligned along the
direction of the cube edges (see Supplemental Material for quantitative
details~\cite{SM}).

We now give a rigorous argument for the existence of asymmetric clusters
in drilling percolation. Fix some $p \in (p_{2D}, p_c)$, where $p_{2D}
\approx 0.5927 < p_c$ is the critical threshold for $2D$ site
percolation. Consider a lattice size $L_X \times L_Y \times L_Z$ with
$L_X = L_Y = L$ and $L_Z = e^{L}$ and take a square domain $A$ in its
base with side length $k = \sqrt{c_0 \log(L)}$ where $c_0$ is a positive
constant that is smaller than $-\left[\log{(p(1-p))}\right]^{-1}$. 
Say that the event $\mathcal{S}(A)$ occurred if, 
along the $z$-direction, no point inside $A$ is drilled but all points
on its boundary are. For large $L$, this event happens with probability at
least ${L}^{c_0\log{(p(1-p))} }$.  Consider also two rectangles $R_x$ and
$R_y$ in the $(x,z)$ and $(y,z)$-planes, respectively, aligned with $A$. 
These rectangles have base length $k$ and height $\exp\{c_1 k\}$, where
$c_1$ is an arbitrary positive constant smaller than the correlation length
for two-dimensional percolation with parameter $p$.
 The
event $\mathcal{R}_x$ (respectively $\mathcal{R}_y$) indicates the
existence of a path crossing $R_x$ (respectively $R_y$) from bottom to
top that has not been drilled in the $y$ (respectively $x$) direction.
By our choice of $c_1$, $\mathcal{R}_x$ and $\mathcal{R}_y$ have
positive probability (uniformly over $k$), \textit{i.e.}~there exists a
$\delta>0$ such that $P(\mathcal{R}_x) \geq \delta$. In addition, if
$\mathcal{S}(A)$ occurs, then there exists a cluster spanning $A \times
[0, e^{c_1 k}]$ from bottom to top and whose projection into the
$(x,y)$-plane does not extend beyond $A$. Thus the probability of
finding a cluster of radius $k$ and height $e^{c_1 k}$ is bounded from
below by the probability that there exists a square $A$ along the
diagonal $x=y$ for which $\mathcal{S}(A) \cap \mathcal{R}_X \cap
\mathcal{R}_Y$ occurs, which is greater than
\begin{equation*}
\begin{split}
 & 1- (1- \delta^2 L^{c_0 \log{[p(1-p)]}})^{L(c_0 \log{L})^{-1/2}} \\
& \geq 1- \exp(-c_0^{-1/2}\delta^2 L^{1+c_0 \log{[p(1-p)]}}
\log(L)^{-1/2}) \ \ ,
\end{split}
\end{equation*}
which converges to unity as $L$ increases. This shows that one expects to
have clusters extremely aligned along the $z$ axis, as numerically
observed (see for example
Fig.~S15 of the
Supplemental Material~\cite{SM}). In fact, the same argument can be
straightforwardly extended to explain the alignment along the $x$ and
$y$ directions, as also observed.

\begin{figure}
\includegraphics[width=\columnwidth]{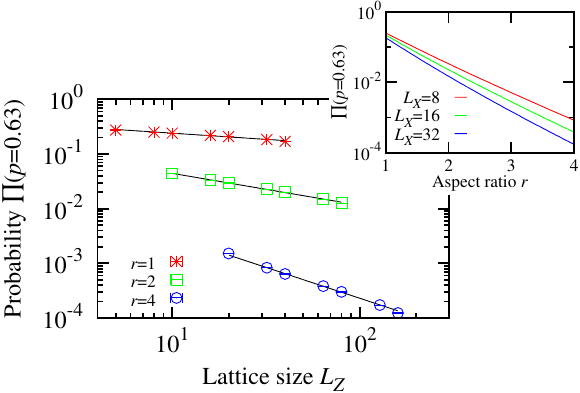}
\caption{
\label{fig::zj_spanning_power_063_inset}
(color online) Main plot: Spanning probability $\Pi$, at ${p=0.63<p_c}$,
as function of the lattice size $L_Z$, for different aspect ratios $r$.
Solid black lines are guides to the eye with slopes, $-0.26$, $-0.58$,
and $-1.22$, for ${r=1}$, $2$, and $4$. The inset shows the same
probability as function of the aspect ratio $r$, for different fixed
lattice sizes $L_X$. Results are based on at least $10^7$ samples.
}
\end{figure}

\textit{Spanning probability.} To understand the properties of the
drilling transition in terms of global connectivity, we consider the
spanning probability $\Pi(p)$, defined as the probability to have at
least one cluster including sites from the top and bottom of the
lattice, at a given value of the control parameter $p$.
Figure~\ref{fig::zj_spanning_power_063_inset} shows the spanning
probability below the threshold $\Pi(p=0.63)$, for different lattice
aspect ratios. The lattice size is ${L_X \times L_Y \times L_Z}$ with
${L_X=L_Y}$ and ${L_Z=rL_X}$ and the spanning probability is measured in
the $z$-direction.  At the drilling transition, $p=p_c$, the spanning
probability approaches a constant for large lattice sizes (see
Supplemental Material,
Fig.~S9), similar to what is
observed for classical percolation~\cite{Cardy92, Smirnov01}.  By
contrast, for values of $p$ between $p_{2D}$ and $p_*$, the numerical results
suggest a power-law decay of the spanning probability with $L_Z$, where
the exponent increases with the aspect ratio $r$. For fixed $L_X$, it
decays exponentially with $r$ (see inset of
Fig.~\ref{fig::zj_spanning_power_063_inset}).

It is possible to establish rigorously the off-critical power-law decay
of $\Pi(p)$, modifying the argument for the existence of anisotropic
clusters presented above. Specifically, we can show that $\Pi(p) \geq
L_X^{-\theta}$, where $\theta=\theta(p,r)>0$, for any fixed $r > 0$ and
$p \in (p_{2D}, p_c)$.  For that, let $B$ be the diagonal band $\{(x,y);
|x - L_X/2| \leq \alpha n \log (L_X/n)$, $|x-y|<2n\}$ in the center of the
$(x,y)$-face of the cube, where $\alpha$ and $n$ are constants setting
the length and width, respectively. Let us say that the event
$\mathcal{B}$ occurred if $B$ is free of holes in the $(x,y)$ plane
\emph{i.e.}~if no sites in $B$ are drilled in the $z$-direction.  Also
say that the event $\mathcal{C}$ occurred if there exists a path
$\sigma$ starting at height $z=0$ and finishing at height $z=L_Z$ whose
projection into the $(x,y)$-plane is contained in $B$ and whose
projection into the $(x,z)$-plane ($(y,z)$-plane) consists of sites
that have not been drilled in the $y$-direction ($x$-direction). As
discussed in detail in the Supplemental Material~\cite{SM}, for well
chosen values of $\alpha$ and $n$ the probability of the event
$\mathcal{C}$ is bounded from below by a constant not depending on $L_X$.
Furthermore $\mathcal{B}$ and $\mathcal{C}$ are independent events.
Since their joint occurrence implies the existence of a cluster
including sites from the bottom and the top of the lattice, we conclude
that
\begin{equation*}
  \begin{split}
    \Pi(p) & \geq P[\mathcal{B} \cap \mathcal{C}] =
P[\mathcal{B}]P[\mathcal{C}]  \\
    & \geq \exp\{- c_2 \alpha\, n \log(L_X)\} c_3 \geq {L_X}^{-\theta} \
\ ,
  \end{split}
\end{equation*}
where $c_2$, $c_3$, and $\theta$ are positive constants that depend on
$p$.

The above argument also shows the existence of anisotropic clusters,
sharpening the numerical results presented before.  For $p < p_{2D}$,
one has $\Pi(p) \sim e^{-c_4 L_{X}}$, similarly to what happens for
uncorrelated random percolation, where $c_4$ also depends on $p$. This
is due to the fact that the projection of a path spanning the lattice
into at least one of the coordinate planes is a path that spans the
corresponding face, which has exponentially small probability in $L_X$,
due to the classical exponential decay of connectivity in the
subcritical phase~\cite{Menshikov86,Aizenman87}.


\emph{Conclusion.} We find unexpected critical behavior when
sequentially drilling holes through a solid cube until it is completely
fragmented. At the critical density of drilled holes, a continuous
transition is observed in a different universality class than the one of
random percolation. We also numerically observe off-critical scale-free
behavior that we can justify for a wide range of densities of holes using
rigorous arguments. 
This model is a representative of more complex percolation models where
sites are removed in a strongly correlated 
manner~\cite{Araujo14, Hilario15, Liu15}. Examples are models where the
set of removed sites is given by randomized trajectories, such as the
so called Pacman and interlacement percolation models proposed to
study the relaxation at the glass transition~\cite{Pastore13}, enzyme gel
degradation~\cite{Abete04} and corrosion~\cite{Sznitman09, Sznitman10}, 
as well as percolation models for distributed 
computation~\cite{Coppersmith93, Winkler00}. Other examples are 
percolation models explicitly introduce strong directional correlations 
as in the removal of cylinders~\cite{Tykesson12} and different variants 
of the four-vertex model \cite{Pete08}.
It would be interesting to explore up to which degree these
models are in the same universality class or share common features.

While the fractal dimension of the largest fragment is consistent with
the one of random percolation, all the other critical exponents are
different. This has practical implications as the connectivity and
transport properties do change considerably close to the threshold of
connectivity. For example, we find the exponent of the order parameter
to be substantially larger than for usual percolation which implies that
the drilling transition is less abrupt. Since sites are removed along a
line, it is necessary to remove more sites to produce the same effect in
the largest fragment. We also find that, compared to usual percolation,
the fractal dimension of the backbone is larger and the one of the
shortest path is smaller, corresponding to a more compact backbone and
therefore enhanced conductivity properties.

%
\begin{acknowledgments}
We acknowledge financial support from the ETH Risk Center, the Brazilian
institute INCT-SC, and grant number FP7-319968 of the European Research
Council.  KJS acknowledges support by the Swiss National Science
Foundation under Grants nos. P2EZP2-152188 and P300P2-161078. MH was
supported by CNPq grant 248718/2013-4 and by ERC AG ``COMPASP''. NA
acknowledges financial support from the Portuguese Foundation for
Science and Technology (FCT) under Contracts nos.
EXCL/FIS-NAN/0083/2012, UID/FIS/00618/2013, and IF/00255/2013.  AT is
grateful for the financial support from CNPq, grants 306348/2012-8 and
478577/2012-5. Part of the rigorous arguments presented in this work were
based on mathematical results developed in the PhD Thesis of one of the
authors, MRH~\cite{Hilario11}.
\end{acknowledgments}
\bibliography{bibliography.bib}

\includepdf[pages={1,2}]{drilling_SM.pdf}
\includepdf[pages={1,3}]{drilling_SM.pdf}
\includepdf[pages={1,4}]{drilling_SM.pdf}
\includepdf[pages={1,5}]{drilling_SM.pdf}
\includepdf[pages={1,6}]{drilling_SM.pdf}
\includepdf[pages={1,7}]{drilling_SM.pdf}
\includepdf[pages={1,8}]{drilling_SM.pdf}
\includepdf[pages={1,9}]{drilling_SM.pdf}

\end{document}